\documentclass[review]{elsarticle}

\usepackage{lineno,hyperref}
\usepackage{amsmath}
\usepackage{amssymb}
\usepackage{algorithmic}
\usepackage{algorithm}
\usepackage{caption}
\usepackage{mathtools}
\usepackage{changes}
\usepackage{multirow}
\usepackage{verbatim}
\usepackage{mathrsfs}
\usepackage{graphicx}
\usepackage{amsmath,amsthm,bm,mathrsfs}

\theoremstyle{plain}% Theorem-like structures provided by amsthm.sty

\theoremstyle{definition}

\theoremstyle{remark}

\modulolinenumbers[5]

\journal{ArXiv.org}

\begin{document}

\begin{frontmatter}

\title{Data-driven $\mathcal{H}_2$-optimal Model Reduction via Offline Transfer Function Sampling}

\author[uz]{Umair~Zulfiqar\corref{mycorrespondingauthor}}
\cortext[mycorrespondingauthor]{Corresponding author}
\ead{umairzulfiqar@shu.edu.cn}
\address[uz]{School of Electronic Information and Electrical Engineering, Yangtze University, Jingzhou, Hubei, 434023, China}
\begin{abstract}
$\mathcal{H}_2$-optimal model order reduction algorithms represent a significant class of techniques, known for their accuracy, which has been extensively validated over the past two decades. Among these, the Iterative Rational Krylov Algorithm (IRKA) is widely regarded as a benchmark for constructing $\mathcal{H}_2$-optimal reduced-order models. However, a key challenge in its data-driven implementation lies in the need for transfer function samples and their derivatives, which must be updated iteratively. Conducting new experiments to acquire these samples each time IRKA updates the interpolation data is impractical. Additionally, for discrete-time systems, obtaining transfer function samples at frequencies outside the unit circle is challenging, as these are not easily accessible through measurements. This paper proposes a method to sample the transfer function and its derivative offline using frequency or time-domain data, which is commonly measured for various design and analysis purposes in industry. By leveraging this approach, there is no need to directly measure transfer function samples at interpolation points, as these can be generated offline using the pre-existing data. This facilitates the offline implementation of IRKA within the frequency- or time-domain Loewner framework. The approach is also extended to discrete-time systems in this work. A numerical example is provided to validate the theoretical findings presented.
\end{abstract}


\end{frontmatter}

%\linenumbers
\section{Introduction}
Consider an $n^{th}$-order $p\times m$ transfer function $G(s)$ of a continuous-time linear time-invariant (LTI) system with $m$ inputs and $p$ outputs. The state-space realization $(A, B, C)$ is related to the transfer function $G(s)$ by the expression:
\begin{align}
G(s)&=C(sI-A)^{-1}B,\nonumber
\end{align}where $A \in \mathbb{R}^{n \times n}$, $B \in \mathbb{R}^{n \times m}$, and $C \in \mathbb{R}^{p \times n}$. The impulse response $h(t)$ of $G(s)$ is given by $h(t) = Ce^{At}B$.

The $r^{th}$-order reduced-order model (ROM) $\hat{G}(s)$ derived through Petrov-Galerkin projection is expressed as:
\begin{align}
\hat{G}(s)&=\hat{C}(sI-\hat{A})^{-1}\hat{B},\nonumber
\end{align}where $\hat{A} = \hat{W}^T A \hat{V} \in \mathbb{R}^{r \times r}$, $\hat{B} = \hat{W}^T B \in \mathbb{R}^{r \times m}$, and $\hat{C} = C \hat{V} \in \mathbb{R}^{p \times r}$. The oblique projector $\Pi = \hat{V} \hat{W}^T$ satisfies the Petrov-Galerkin projection condition $\hat{W}^T \hat{V} = I$.

Let $T \in \mathbb{C}^{r \times r}$ be an invertible matrix. The projection matrices $\hat{W}$ and/or $\hat{V}$ can be replaced with $\hat{W}T$ and/or $\hat{V}T$, resulting in the same ROM $\hat{G}(s)$ but with a different state-space realization. This property is leveraged to convert complex projection matrices into real ones. Consequently, for the remainder of this paper, we will assume $\hat{V}$ and $\hat{W}$ to be complex matrices without loss of generality. Additionally, this property is used to enforce the Petrov-Galerkin projection condition $\hat{W}^*\hat{V} = I$ by setting  $T=(\hat{W}^*\hat{V})^{-1}$ or $T=(\hat{V}^*\hat{W})^{-1}$. Throughout this paper, we assume that $\hat{W}^*\hat{V}$ is invertible due to the full column rank of $\hat{V}$ and $\hat{W}$. Moreover, $A$ and $\hat{A}$ are assumed to be Hurwitz.
\subsection{Review of Interpolation Theory}
Given the right interpolation points $\sigma_1,\sigma_2,\cdots,\sigma_r$ and left interpolation points $\mu_1,\mu_2,\cdots,\mu_r$ along with the corresponding right tangential directions $b_1,b_2,\cdots,b_r$ and left tangential directions $c_1,c_2,\cdots,c_3$, we construct the projection matrices $\hat{V}\in\mathbb{C}^{n\times r}$ and $\hat{W}\in\mathbb{C}^{n\times r}$ as follows:
\begin{align}
\hat{V}&=\begin{bmatrix}(\sigma_1I-A)^{-1}Bb_1&\cdots&(\sigma_rI-A)^{-1}Bb_r\end{bmatrix},\label{Vr}\\
\hat{W}^*&=\begin{bmatrix}c_1C(\mu_1I-A)^{-1}\\\vdots\\c_rC(\mu_rI-A)^{-1}\end{bmatrix}.\label{Wr}
\end{align}
The ROM obtained using the oblique projector $\Pi=\hat{V}(\hat{W}^*\hat{V})^{-1}\hat{W}^*$ satisfies the following interpolation conditions:
\begin{align}
G(\sigma_i)b_i&=\hat{G}(\sigma_i)b_i,\label{int1}\\
c_iG(\mu_i)&=c_i\hat{G}(\mu_i),\label{int2}
\end{align}for $i=1,\cdots,r$. Moreover, when $\sigma_i=\mu_i$, the following Hermite interpolation conditions are also satisfied:
\begin{align}
c_iG^{\prime}(\sigma_i)b_i&=\hat{G}^{\prime}(\sigma_i)b_i,\nonumber
\end{align}for $i=1,\cdots,r$.

We now introduce the data-driven construction of an interpolation framework that utilizes frequency domain samples of $G(s)$ to build the ROM. In this framework, the Loewner matrix $\mathbb{L}=\hat{W}^*\hat{V}$ is defined as follows:
\begin{align}
\mathbb{L}=\begin{bmatrix}-\frac{c_1\big(G(\sigma_1)-G(\mu_1)\big)b_1}{\sigma_1-\mu_1}&\cdots&-\frac{c_1\big(G(\sigma_r)-G(\mu_1)\big)b_r}{\sigma_r-\mu_1}\\
-\frac{c_r\big(G(\sigma_1)-G(\mu_r)\big)b_1}{\sigma_1-\mu_r}&\cdots&-\frac{c_r\big(G(\sigma_r)-G(\mu_r)\big)b_r}{\sigma_r-\mu_r}
\end{bmatrix}.\nonumber
\end{align}When $\sigma_i=\mu_i$, the term $\frac{G(\sigma_j)-G(\mu_i)}{\sigma_j-\mu_i}$ is replaced with $G^{\prime}(\sigma_i)$ in $\mathbb{L}$. Additionally, the shifted Loewner matrix $\mathbb{L}_s=\hat{W}^*A\hat{V}$ is given by:
\begin{align}
\mathbb{L}_s=\begin{bmatrix}-\frac{c_1\big(\sigma_1G(\sigma_1)-\mu_1G(\mu_1)\big)b_1}{\sigma_1-\mu_1}&\cdots&-\frac{c_1\big(\sigma_rG(\sigma_r)-\mu_1G(\mu_1)\big)b_r}{\sigma_r-\mu_1}\\
-\frac{c_r\big(\sigma_1G(\sigma_1)-\mu_rG(\mu_r)\big)b_1}{\sigma_1-\mu_r}&\cdots&-\frac{c_r\big(\sigma_rG(\sigma_r)-\mu_rG(\mu_r)\big)b_r}{\sigma_r-\mu_r}
\end{bmatrix}.\nonumber
\end{align}Here too, when $\sigma_i=\mu_i$, $\frac{G(\sigma_j)-G(\mu_i)}{\sigma_j-\mu_i}$ is replaced with $G^{\prime}(\sigma_i)$ in $\mathbb{L}_s$. The matrices $\hat{B}$ and $\hat{C}$ contain samples of $G(s)$ at $\sigma_i$ and $\mu_i$ in the directions $b_i$ and $c_i$, respectively:
\begin{align}
\hat{B}&=\mathbb{L}^{-1}\begin{bmatrix}c_1G(\mu_1)\\\vdots\\c_rG(\mu_r)\end{bmatrix},&&&\hat{C}&=\begin{bmatrix}G(\sigma_1)b_1&\cdots&G(\sigma_r)b_r\end{bmatrix}.\nonumber
\end{align}Finally, the ROM that satisfies the interpolation conditions (\ref{int1}) and (\ref{int2}) is obtained by setting $\hat{A}=\mathbb{L}^{-1}\mathbb{L}_s$. Due to the role of the Loewner and shifted Loewner matrices in this data-driven construction, this approach is commonly referred to as the Loewner framework (LF) in the literature \citep{gosea2022data}.
\subsection{$\mathcal{H}_2$-optimal MOR \citep{gugercin2008h_2}}
Assume that $G(s)$ and $\hat{G}(s)$  have simple poles. Then, they can be expressed in the following pole-residue form:
\begin{align}
G(s)&=\sum_{k=1}^{n}\frac{l_kr_k^*}{s-\lambda_k},&&&\hat{G}(s)&=\sum_{k=1}^{r}\frac{\hat{l}_k\hat{r}_k^*}{s-\hat{\lambda}_k}.\nonumber
\end{align}
The necessary conditions for a local optimum of $||G(s)-\hat{G}(s)||_{\mathcal{H}_2}^2$ are as follows:
\begin{align}
\hat{l}_i^*G^{\prime}(-\hat{\lambda}_i)\hat{r}_i&=\hat{l}_i^*\hat{G}^{\prime}(-\hat{\lambda}_i)\hat{r}_i,\label{op1}\\
\hat{l}_i^*G(-\hat{\lambda}_i)&=\hat{l}_i^*\hat{G}(-\hat{\lambda}_i),\label{op2}\\
G(-\hat{\lambda}_i)\hat{r}_i&=\hat{G}(-\hat{\lambda}_i)\hat{r}_i,\label{op3}
\end{align}for $i=1,2,\cdots,r$. These Hermite interpolation conditions can be satisfied by setting $\sigma_i=-\hat{\lambda}_i$, $b_i=\hat{r}_i$, and $c_i=\hat{l}_i^*$ in the data-driven interpolation framework, assuming that we have access to the samples $G(-\hat{\lambda_i})\hat{r}_i$, $\hat{l}_i^*G(-\lambda_i)$, and $G^{\prime}(-\hat{\lambda_i})$. However, since the ROM, along with its poles and residues, is initially unknown, an iterative correction process is required, starting with an arbitrary guess of the poles and residues of $\hat{G}(s)$.

This iterative approach is central to the Iterative Rational Krylov Algorithm (IRKA), which is a leading method for $\mathcal{H}_2$-optimal model order reduction (MOR). Upon convergence, IRKA constructs a local optimum of $||G(s)-\hat{G}(s)||_{\mathcal{H}_2}^2$ and typically achieves this quickly, even when initialized with arbitrary interpolation data.

In a data-driven context, however, the challenge lies in the need to halt the algorithm each time the interpolation data is updated, to conduct new experiments and measure $G(\sigma_i)b_i$, $c_iG(\sigma_i)$, and $G^{\prime}(\sigma_i)$. This requirement for repeated experiments makes the online application of IRKA impractical, as it demands significant time, effort, and resources. This limitation motivates the development of new methods capable of constructing $G(\sigma_i)b_i$, $c_iG(\sigma_i)$ and $G^{\prime}(\sigma_i)$ from existing offline time- and frequency-domain data, which is the primary focus of this paper.
\section{Main Work}
In this section, we will introduce both frequency- and time-domain methods for generating new samples of $G(s)$ from existing data, enabling the offline implementation of IRKA in a data-driven context.
\subsection{Data-driven $\mathcal{H}_2$-suboptimal MOR}
Before presenting the main results, we will first discuss an algorithm known in the literature as the ``Pseudo-optimal Rational Krylov (PORK)" algorithm \citep{wolf2014h}. This algorithm constructs a ROM in a single run that satisfies either the optimality condition (\ref{op2}) or (\ref{op3}). Notably, PORK can be implemented offline in a data-driven setting without any modifications. Let us define $S=diag(\sigma_1,\cdots,\sigma_r)$ and $L_b=\begin{bmatrix}b_1,\cdots,b_r\end{bmatrix}$. The observability Gramian $Q_s$ of the pair $(-S,L_b)$ solves the following Lyapunov equation:
\begin{align}
-S^*Q_s+Q_sS+L_b^*L_b&=0.\nonumber
\end{align}If the interpolation points $\sigma_1,\cdots,\sigma_r$ have positive real parts and the pair $(-S,L_b)$ is observable, then the optimality condition condition (\ref{op3}) can be satisfied by implicitly enforcing the Petrov-Galerkin projection condition $\hat{W}^*\hat{V}=I$ without explicitly constructing $\hat{W}$. The following ROM satisfies the optimality condition (\ref{op3}):
\begin{align}
\hat{A}&=-Q_s^{-1}S^*Q_s,&\hat{B}&=-Q_s^{-1}L_b^*,&\hat{C}&=C\hat{V}=\begin{bmatrix}G(\sigma_1)b_1&\cdots&G(\sigma_r)b_r\end{bmatrix}.\nonumber
\end{align}

Dually, the optimality condition (\ref{op2}) can be satisfied by implicitly enforcing the Petrov-Galerkin projection condition $\hat{W}^*\hat{V}=I$ without explicitly constructing $\hat{V}$. Define $U=diag(\mu_1,\cdots,\mu_r)$ and $L_c=\begin{bmatrix}c_1^*&\cdots&c_r^*\end{bmatrix}^*$. The controllability Gramian $P_s$ for the pair $(-U,L_c)$ satisfies the following Lyapunov equation:
\begin{align}
-UP_s+P_sU^*+L_cL_c^*&=0.\nonumber
\end{align}Assuming that the interpolation points $\mu_1,\cdots,\mu_r$ have positive real parts and the pair $(-U,L_c)$ is controllable, a ROM that satisfies the optimality condition (\ref{op2}) can be derived as follows:
\begin{align}
\hat{A}&=-P_sU^*P_s^{-1},&\hat{B}&=\hat{W}^*B=\begin{bmatrix}c_1G(\mu_1)\\\vdots\\c_rG(\mu_r)\end{bmatrix},&\hat{C}&=-L_c^*P_s^{-1}.\nonumber
\end{align}

It is now evident that if samples of $G(s)$ are available at the interpolation points, the optimality conditions (\ref{op2}) or (\ref{op3}) can be satisfied in a single run. Additionally, these optimality conditions possess the property that as the number of interpolation points increases, the error $||G(s)-\hat{G}(s)||_{\mathcal{H}_2}$ decreases monotonically. However, this does not indicate how quickly the error decays, meaning the accuracy is highly dependent on the interpolation data. The advantage of IRKA lies in its iterative refinement of the interpolation data, which enhances accuracy. IRKA can achieve high accuracy even if it does not fully converge. In practice, IRKA typically provides a very accurate ROM after only a few iterations, even if it does not reach the local optimum. Therefore, while PORK can serve as an alternative for $\mathcal{H}_2$-MOR in a data-driven setting, it cannot match the accuracy of IRKA with arbitrary interpolation data.
\subsection{Offline Sampling of $G(s)$ using Available Frequency-domain Data}
In industries such as aerospace, defense, and automotive, frequency domain data is collected to construct the Fourier transform $G(j\omega)$ by exciting systems at various frequencies $\omega$ rad/sec. This data is crucial for various analysis and design tasks, including system identification, control design, resonance frequency calculation, and vibration analysis, among others \citep{lennart1999system, ozbay2018frequency, pintelon2008frequency, gillberg2006frequency, morelli2020practical}. Given the reliance of many design and analysis procedures on this data, it is essential for the safe and effective operation of systems to construct the Fourier transform. In this subsection, we demonstrate that this existing data is sufficient to generate offline samples of $G(s)$  at any desired complex frequency $s=\sigma+j\omega$.

The projection matrices $\hat{V}$ and $\hat{W}$, defined in (\ref{Vr}) and (\ref{Wr}) respectively, satisfy the following Sylvester equations:
\begin{align}
A\hat{V}-\hat{V}S+BL_b&=0,\nonumber\\
-U\hat{W}^*+\hat{W}^*A+L_cC&=0,\nonumber
\end{align}as discussed in \cite{wolf2014h,wolf2013h}. When the interpolation points $\sigma_i$ and $\mu_i$ have positive real parts, $\hat{V}$ and $\hat{W}$ can be computed using the integral expressions:
\begin{align}
\hat{V}&=\frac{1}{2\pi}\int_{-\infty}^{\infty}(j\omega I-A)^{-1}BL_b(-j\omega I+S)^{-1}d\omega,\nonumber\\
\hat{W}^*&=\frac{1}{2\pi}\int_{-\infty}^{\infty}(-j\omega I+U)^{-1}L_cC(j\omega I-A)^{-1}d\omega,\nonumber
\end{align}cf. \citep{sorensen2002sylvester}. Consequently,
\begin{align}
\begin{bmatrix}G(\sigma_1)b_1&\cdots&G(\sigma_r)b_r\end{bmatrix}&=\frac{1}{2\pi}\int_{-\infty}^{\infty}G(j\omega)L_b(-j\omega I+S)^{-1}d\omega,\nonumber\\
\begin{bmatrix}c_1G(\mu_1)\\\vdots\\c_rG(\mu_r)\end{bmatrix}&=\frac{1}{2\pi}\int_{-\infty}^{\infty}(-j\omega I+U)^{-1}L_cG(j\omega)d\omega.\nonumber
\end{align}

These integrals can be approximated using numerical quadrature methods. For simplicity, we employ the Trapezoidal rule with uniform sampling in this paper, although adaptive quadrature methods can offer improved accuracy. With $\Delta\omega = \omega_k - \omega_{k-1}$ and $\omega_0 = -\infty$, the approximations are given by:
\begin{align}
\begin{bmatrix}G(\sigma_1)b_1&\cdots&G(\sigma_r)b_r\end{bmatrix}&\approx\frac{\Delta \omega}{4\pi}\sum_{k=1}^{\infty}\Big(T_\sigma(\omega_{k-1})+T_\sigma(\omega_k)\Big),\nonumber\\
\begin{bmatrix}c_1G(\mu_1)\\\vdots\\c_rG(\mu_r)\end{bmatrix}&\approx\frac{\Delta \omega}{4\pi}\sum_{k=1}^{\infty}\Big(T_\mu(\omega_{k-1})+T_\mu(\omega_k)\Big);\nonumber
\end{align}where
\begin{align}
T_\sigma(\omega)&=G(j\omega)L_b(-j\omega I+S)^{-1},\nonumber\\
T_\mu(\omega)&=(-j\omega I+U)^{-1}L_cG(j\omega).\nonumber
\end{align}

Theoretically, infinite samples of $G(j\omega)$ are required to compute these integrals. However, in practice, the bandwidths of most practical dynamical systems are limited, and they do not operate over the entire frequency range. Beyond the system's actual bandwidth, $G(j\omega)$ is essentially zero. Thus, it suffices to sample $G(j\omega_k)$ at positive frequencies within the system's bandwidth, since $G(-j\omega) = G^*(j\omega)$. While reviewing methods for obtaining such frequency domain data is beyond this paper's scope, it is worth noting that a single output measurement to a chirp input with rapidly varying frequency can effectively capture the frequency response of $G(s)$ within its bandwidth \citep{pintelon2012system}.

It is interesting to note that to compute these numerical integration, we do not need to sample $G(s)$ at $\sigma_i$ or $\mu_i$. Therefore, if the frequency domain data $G(j\omega_k)$ within the bandwidth of $G(s)$ is available, we can sample $G(s)$ offline at any complex frequency $s_k=\sigma_k+j\omega_k$ with positive real part. Consequently, we can run IRKA offline with pre-computed frequency domain data $G(j\omega_k)$. Moreover, we can restart IRKA with any arbitrary interpolation data without at problem since we do not need any online computation once we have obtained the frequency domain data within the bandwidth of original system.

To estimate the values of $G^{\prime}(s_i)$,  numerical differentiation can be employed. By slightly perturbing $s_i$ to $s_i + \Delta s$, one can approximate $G^{\prime}(s_i)$ by computing the slope:
\begin{align}
G^{\prime}(s_i)\approx \frac{G(s_i+\Delta s)-G(s_i)}{\Delta s}.\nonumber
\end{align}
This method will be referred to as the ``Frequency-domain Quadrature-based Loewner Framework (FQLF)" in this paper.
\subsubsection{Example 1}
Consider a $6^{th}$-order system with $3$ inputs and $2$ outputs characterized by the following state-space realization:
\begin{align}
A &=\begin{bsmallmatrix}
2.0405&-4.0202&3.6597&-3.0030&-1.3991&0.0223\\
10.7565&-2.6355&3.0279&1.1415&-2.8439&-11.4753\\
0&0&-5.0713&-0.7836&3.5921&5.3577\\
0&0&0&-0.7377&0.4368&2.0851\\
0&0&0&0&-2.4419&-0.5944\\
0&0&0&0&0&-1.9241
\end{bsmallmatrix},\nonumber\\
B&=\begin{bsmallmatrix}-1.9113&0.5763&1.6860\\
    -1.3247&4.8135&9.5314\\
     1.9909&-3.7210&-5.1055\\
    -0.9923&1.5222&-1.5323\\
    -0.5533&1.9860&0.7948\\
     1.0127&1.3013&3.4008
     \end{bsmallmatrix},\nonumber\\
C&=\begin{bsmallmatrix}0.0680&0.2673&0.0594&0.4300&-0.0130&-0.1340\\
    0.1494&-0.1304&0.2256&0.5877&-0.0828&0.1467\end{bsmallmatrix}.\nonumber
\end{align}The following interpolation data is used in this example to construct $4^{th}$-order ROMs using LF and FQLF:
\begin{align}
S &= diag(5+j7,5-j7,3+j2,3-j2),\nonumber\\
U&=diag(0.1+j6,0.1-j6,0.5+j1,0.5-j1),\nonumber\\
L_b&=\begin{bsmallmatrix}1+j2&1-j2&3+j4&3-j4\\
     5+j6&5-j6&7+j8&7-j8\\
     9+j10&9-j10&11+j12&11-j12
     \end{bsmallmatrix},\nonumber\\
L_c&=\begin{bsmallmatrix}13+j14&15+j16\\
    13-j14&15-j16\\
    17+j18&19+j20\\
    17-j18&19-j20
    \end{bsmallmatrix}.\nonumber
\end{align} For the LF, the exact values of $G(\sigma_i)$ and $G(\mu_i)$ are used. For the FQLF, these values are approximated using the Trapezoidal rule from frequency domain data $G(j\omega)$, which spans from $\omega = [0, 500]$ rad/sec with $25,000$ uniformly distributed samples.

The ROM constructed using LF that exactly satisfies the specified interpolation conditions is given by:
\begin{align}
\hat{A}&=\begin{bsmallmatrix}12.8203&15.3845&8.9487&9.5128\\
-21.7995&-11.3167&-17.8338&-19.3510\\
-2.8202&-3.1235&-0.4269&-1.7302\\
-7.4980&-8.2729&-11.0477&-6.8226\end{bsmallmatrix},\nonumber\\
\hat{B}&=\begin{bsmallmatrix}0.0842&0.5174&-1.1657\\
-0.4234&0.5606&1.3800\\
-0.1007&0.1789&0.2252\\
0.2188&-0.5686&1.1247
\end{bsmallmatrix},\nonumber\\
\hat{C}&=\begin{bsmallmatrix}-0.2705&2.1923&-0.9447&-1.6136\\
-1.8891&-2.5726&4.1727&-2.0704\end{bsmallmatrix}.\nonumber
\end{align}The ROM constructed using FQLF is:
\begin{align}
\hat{A}&=\begin{bsmallmatrix}12.7743&15.3316&8.8888&9.4461\\
  -21.7762&-11.2899&-17.8035&-19.3172\\
   -2.8162&-3.1189&-0.4216&-1.7244\\
   -7.4581&-8.2270&-10.9960&-6.7649\end{bsmallmatrix},\nonumber\\
\hat{B}&=\begin{bsmallmatrix}0.0885&0.5128&-1.1585\\
   -0.4256&0.5629&1.3764\\
   -0.1011&0.1793&0.2245\\
    0.2151&-0.5646&1.1185\end{bsmallmatrix},\nonumber\\
\hat{C}&=\begin{bsmallmatrix}-0.2820&2.1796&-0.9585&-1.6287\\
   -1.8727&-2.5541&4.1931&-2.0480\end{bsmallmatrix}.\nonumber
\end{align}Upon inspection, the state-space realizations of these ROMs are found to be quite close to those constructed using LF.

The derivatives $G^{-\prime}(\sigma_1)b_1$ and $G^{-\prime}(\sigma_2)b_2$ are given by:
\begin{align}
G^{-\prime}(\sigma_1)b_1&=\begin{bmatrix} -0.2523 + j0.6019\\
  -0.2884 - j0.5045\end{bmatrix},\nonumber\\
G^{-\prime}(\sigma_3)b_3&=\begin{bmatrix} 0.3699 - j1.1534\\
   2.3440 - j0.7540\end{bmatrix}.\nonumber
\end{align}
Next, using $\Delta s=10^{-4}(1+j1)$, the derivatives $G^{-\prime}(\sigma_1)b_1$ and $G^{-\prime}(\sigma_3)b_3$ estimated with FQLF is given by:
\begin{align}
G^{-\prime}(\sigma_1)b_1&\approx\begin{bmatrix} -0.2522 + j0.6019\\
  -0.2884 - j0.5045\end{bmatrix},\nonumber\\
G^{-\prime}(\sigma_3)b_3&\approx\begin{bmatrix} 0.3698 - j1.1534\\
   2.3439 - j0.7540\end{bmatrix}.\nonumber
\end{align}
These estimates are quite close to the actual values.
\subsection{Offline Sampling of $G(s)$ using Available Time-domain Data}
In many applications, making frequency domain measurements is not practical. Instead, impulse response data is often used for various analysis and design purposes. Frequency response data in these cases is derived from the impulse response data. When direct impulse response measurements are not feasible, a step input can be used, and the impulse response can be obtained by differentiation. Since a detailed review of these methods is beyond the scope of this paper, we direct readers to \citep{stan2002comparison, finno1998impulse, holters2009impulse, foster1986impulse, borish1983efficient} for more information. In this subsection, we demonstrate that existing impulse response data is sufficient to generate offline samples of  $G(s)$ at any desired complex frequency $s=\sigma+j\omega$.

If the interpolation points $\sigma_i$ and $\mu_i$ have positive real parts, $\hat{V}$ and $\hat{W}$ can be calculated using the following integral expressions:
\begin{align}
\hat{V}&=\int_{0}^{\infty}e^{At}BL_be^{-St}dt,\nonumber\\
\hat{W}^*&=\int_{0}^{\infty}e^{-Ut}L_cCe^{At}dt.\nonumber
\end{align}It follows that
\begin{align}
\begin{bmatrix}G(\sigma_1)b_1&\cdots&G(\sigma_r)b_r\end{bmatrix}&=\int_{0}^{\infty}h(t)L_be^{-St}dt,\nonumber\\
\begin{bmatrix}c_1G(\mu_1)\\\vdots\\c_rG(\mu_r)\end{bmatrix}&=\int_{0}^{\infty}e^{-Ut}L_ch(t)dt.\nonumber
\end{align}These integrals can be approximated using any quadrature rule, such as an adaptive quadrature rule. For simplicity, this paper uses the Trapezoidal rule with uniform sampling, although adaptive quadrature rules can offer better accuracy. By defining $\Delta t=t_k-t_{k-1}$ and $t_0=0$, we obtain:
\begin{align}
\begin{bmatrix}G(\sigma_1)b_1&\cdots&G(\sigma_r)b_r\end{bmatrix}&\approx\frac{\Delta t}{2}\sum_{k=1}^{\infty}\Big(K_\sigma(t_{k-1})+K_\sigma(t_k)\Big),\nonumber\\
\begin{bmatrix}c_1G(\mu_1)\\\vdots\\c_rG(\mu_r)\end{bmatrix}&\approx\frac{\Delta t}{2}\sum_{k=1}^{\infty}\Big(K_\mu(t_{k-1})+K_t(\omega_k)\Big);\nonumber
\end{align}where
\begin{align}
K_\sigma(t)&=h(t)L_be^{-St},\nonumber\\
K_\mu(t)&=e^{-Ut}L_ch(t).\nonumber
\end{align} Theoretically, infinite samples of $h(t)$ are needed to compute these integrals. However, in practice, the impulse response of stable systems decays over time due to the decaying exponential $e^{At}$ in the impulse response expression. The farther the system poles are from the $j\omega$-axis in the left half of the $s$-plane, the faster the impulse response decays. Compared to the frequency response, the impulse response decays more quickly and eventually approaches zero. Therefore, only samples of the impulse response during the time it has not significantly decayed are needed. Thus, if impulse response data is available, $G(s)$ can be sampled offline at any complex frequency $s_k=\sigma_k+j\omega_k$ with a positive real part. Consequently, IRKA can be implemented offline using pre-computed impulse response data. The method described in this subsection will be referred to as the ``Time-domain Quadrature-based Loewner Framework (TQLF)."
\subsubsection{Example 2}
Consider the same system and interpolation data as presented in \textit{Example 1}. For TQLF, $G(\sigma_i)$ and $G(\mu_i)$ are approximated using the Trapezoidal rule from impulse response data $h(t)$, which spans from $t=[0,30]$ seconds and contains $10000$ uniformly distributed samples. The ROM constructed using TQLF is given by:
\begin{align}
\hat{A}&=\begin{bsmallmatrix}12.8218&15.3862&8.9506&9.5149\\
  -21.8003&-11.3175&-17.8348&-19.3520\\
   -2.8204&-3.1237&-0.4271&-1.7305\\
   -7.4992&-8.2742&-11.0493&-6.8244\end{bsmallmatrix},\nonumber\\
\hat{B}&=\begin{bsmallmatrix} 0.0841&0.5173&-1.1659\\
   -0.4234&0.5606&1.3801\\
   -0.1007&0.1789&0.2252\\
    0.2188&-0.5686&1.1248\end{bsmallmatrix},\nonumber\\
\hat{C}&=\begin{bsmallmatrix} -0.2704&2.1926&-0.9445&-1.6134\\
   -1.8892&-2.5730&4.1727&-2.0708
\end{bsmallmatrix}.\nonumber
\end{align}The state-space realization of this ROMs is quite close to those constructed by LF. Furthermore, by setting $\Delta s=10^{-4}(1+j1)$, the estimated values of $G^{-\prime}(\sigma_1)b_1$ and $G^{-\prime}(\sigma_3)b_3$ using TQLF are:
\begin{align}
G^{-\prime}(\sigma_1)b_1&\approx\begin{bmatrix} -0.2522 + 0.6019i\\
  -0.2884 - 0.5045i\end{bmatrix},\nonumber\\
G^{-\prime}(\sigma_3)b_3&\approx\begin{bmatrix} 0.3698 - 1.1534i\\
   2.3440 - 0.7541i\end{bmatrix}.\nonumber
\end{align}
Once again, inspection shows that these estimates are nearly identical to the actual values.
\subsection{Tracking the Error $||G(s)-\hat{G}(s)||_{\mathcal{H}_2}$}
The $\mathcal{H}_2$-norm of the error is expressed as follows:
\begin{align}
||G(s)-\hat{G}(s)||_{\mathcal{H}_2}^2&=||G(s)||_{\mathcal{H}_2}^2-2trace\Big(\hat{C}\begin{bmatrix}b_1^*G(\sigma_1^*)\\\vdots\\b_r^*G(\sigma_r^*)\end{bmatrix}\Big)+||\hat{G}(s)||_{\mathcal{H}_2}^2,\nonumber\\
&=||G(s)||_{\mathcal{H}_2}^2-2trace\Big(\hat{B}^*\begin{bmatrix}cG(\mu_1)\\\vdots\\c_kG(\mu_r)\end{bmatrix}\Big)+||\hat{G}(s)||_{\mathcal{H}_2}^2,\nonumber
\end{align}By monitoring the growth in the following terms:
\begin{align}
2trace\Big(\begin{bmatrix}G(\sigma_1)b_1&\cdots&G(\sigma_r)b_r\end{bmatrix}\hat{C}^*\Big)-||\hat{G}(s)||_{\mathcal{H}_2}^2\nonumber
\end{align}or
\begin{align}
2trace\Big(\hat{B}^*\begin{bmatrix}cG(\mu_1)\\\vdots\\c_kG(\mu_r)\end{bmatrix}\Big)-||\hat{G}(s)||_{\mathcal{H}_2}^2,\nonumber
\end{align}we can track the reduction in $||G(s)-\hat{G}(s)||_{\mathcal{H}_2}^2$ during each iteration of IRKA. The algorithm can be stopped if the growth in these terms levels off, indicating that further refinement of the interpolation data is no longer improving the accuracy of the ROM. Importantly, no additional information is required to monitor the error, as the relevant data for these calculations is already available from the ROM construction process.
\subsection{Extension to Discrete-time Case}
We now extend our approach to the discrete-time case. Let $G(z)$ be a stable system with the state-space realization given by:
\begin{align}
G(z)=C(zI-A)^{-1}B.\nonumber
\end{align}where $z=e^{j\omega}$. The $r^{th}$-order ROM $\hat{G}(z)$ can be represented as:
\begin{align}
\hat{G}(z)=\hat{C}(zI-\hat{A})^{-1}\hat{B}.\nonumber
\end{align}It is assumed that $\hat{G}(z)$ is stable.

Let us define $\mathcal{S}$ and $\mathcal{U}$ as $\mathcal{S}=diag\big(\frac{1}{\sigma_1},\cdots,\frac{1}{\sigma_r}\big)$ and $\mathcal{U}=diag\big(\frac{1}{\mu_1},\cdots,\frac{1}{\mu_r}\big)$. Let $\mathcal{W}$ and $\mathcal{V}$ satisfy the following discrete-time Sylvester equations:
\begin{align}
A\mathcal{V}S-\mathcal{V}+BL_b&=0,\nonumber\\
U\mathcal{W}^*A-\mathcal{W}^*+L_cC&=0.\nonumber
\end{align}
When eigenvalues of $\mathcal{S}$ and $\mathcal{U}$ lie within the unit circle, $\mathcal{V}$ and $\mathcal{W}^*$ can be computed using the following integral expressions:
\begin{align}
\mathcal{V}&=\frac{1}{2\pi}\int_{-\pi}^{\pi}(e^{j\omega}I-A)^{-1}BL_b(e^{-j\omega}I-\mathcal{S})^{-1}d\omega,\nonumber\\
\mathcal{W}^*&=\frac{1}{2\pi}\int_{-\pi}^{\pi}(e^{-j\omega}I-\mathcal{U})^{-1}L_cC(e^{j\omega}I-A)^{-1}d\omega.\nonumber
\end{align}Furthermore,
\begin{align}
C\mathcal{V}\mathcal{S}&=\Big(\frac{1}{2\pi}\int_{-\pi}^{\pi}G(e^{j\omega})L_b(e^{-j\omega}I-\mathcal{S})^{-1}d\omega\Big)\mathcal{S}\nonumber\\
&=\begin{bmatrix}G(\sigma_1)b_1&\cdots&G(\sigma_r)b_r\end{bmatrix}\nonumber\\
&\approx\frac{\Delta \omega}{4\pi}\sum_{k=1}^{N}\Big(\mathcal{T}_\omega(\omega_{k-1})+\mathcal{T}_\sigma(\omega_k)\Big)\mathcal{S},\nonumber\\
\mathcal{U}\mathcal{W}^*B&=\mathcal{U}\Big(\frac{1}{2\pi}\int_{-\pi}^{\pi}(e^{-j\omega}I-\mathcal{U})^{-1}L_cG(e^{j\omega})\Big)\nonumber\\
&=\begin{bmatrix}c_1G(\mu_1)\\\vdots\\c_rG(\mu_r)\end{bmatrix}\nonumber\\
&\approx\mathcal{U}\Big(\frac{\Delta \omega}{4\pi}\sum_{k=1}^{N}\Big(\mathcal{T}_\mu(\omega_{k-1})+\mathcal{T}_\mu(\omega_k)\Big)\nonumber
\end{align}where
\begin{align}
\mathcal{T}_\sigma(\omega)&=G(e^{j\omega})L_b(e^{-j\omega} I-\mathcal{S})^{-1},\nonumber\\
\mathcal{T}_\mu(\omega)&=(e^{-j\omega} I-\mathcal{U})^{-1}L_cG(e^{j\omega}).\nonumber
\end{align}Thus, similar to the continuous-time case, we can sample $G(z)$ offline from existing frequency response data at any $z_i=e^{j\omega_i}$ as long as $\frac{1}{z_i}$ is within the unit circle. This allows sampling of $G(z)$ at frequencies outside the unit circle without directly exciting the system at those frequencies. This is particularly valuable in $\mathcal{H}_2$-optimal MOR, where samples at such frequencies are often needed but are not practical to obtain directly due to safety constraints \citep{ackermann2024time}.

Similarly, by using the impulse response $h(k)=CA^kB$ of $G(z)$, we can approximate:
\begin{align}
\begin{bmatrix}G(\sigma_1)b_1&\cdots&G(\sigma_r)b_r\end{bmatrix}&\approx\Big(\sum_{k=0}^{N}h(k)L_b\mathcal{S}^k\Big)\mathcal{S},\nonumber\\
\begin{bmatrix}c_1G(\mu_1)\\\vdots\\c_rG(\mu_r)\end{bmatrix}&\approx\mathcal{U}\Big(\sum_{k=0}^{N}\mathcal{U}^kL_ch(k)\Big).\nonumber
\end{align}Thus, as in the continuous-time case, $G(z)$ can be sampled offline from available impulse response data at any $z_i=e^{j\omega_i}$, provided that $\frac{1}{z_i}$ is within the unit circle.
\section{Conclusion}
In this research, we have addressed the main bottlenecks that hinder the practical implementation of IRKA in data-driven contexts. Specifically, we introduce both frequency- and time-domain Loewner frameworks, which utilize pre-computed frequency response and impulse response data, respectively, to sample the transfer function at desired interpolation points and construct ROM from these offline-generated samples of the transfer function and its derivative. Additionally, we propose a data-driven method to track error throughout the IRKA iterations, which can serve as a stopping criterion for the algorithm. This approach is also applicable to discrete-time systems. The methodology and its features are validated through an illustrative example, which confirms the theoretical results presented in the paper.
%\bibliography{mybibfile}

\end{document}